\documentclass[draft,jgrga]{agu2001}
\usepackage{graphicx}
\def\apj{ApJ}%
\def\apjl{ApJ}%
\def\apjs{ApJS}%
\def\aap{A\&A}%
\def\mnras{MNRAS}%
\def\solphys{Sol.~Phys.}%
\def\nat{Nature}%
\def\grl{Geophys.~Res.~Lett.}%
\authorrunninghead{BALMACEDA ET AL.}
\titlerunninghead{A homogeneous database of sunspot areas }
\authoraddr{L. A. Balmaceda, current affiliation: GACE/IPL -- Universidad de Valencia, PO BOX 22085, 46071 Valencia, Spain (laura.balmaceda@uv.es)}

\begin{document}

%% ------------------------------------------------------------------------ %%
%
%  ENABLE IMAGE DISPLAY WHILE USING DRAFT MODE
%
%% ------------------------------------------------------------------------ %%
%
% Uncomment the following code (as well as \usepackage{graphicx} above)
% if you need to include images in draft mode
%\setkeys{Gin}{draft=false}
%
% PLEASE NOTE: WHEN YOU SUBMIT YOUR LATEX FILE TO GEMS, COMMENT OUT ANY COMMANDS
% THAT INCLUDE GRAPHICS.
% (See FIGURES section near the end of the file)
%

%% ------------------------------------------------------------------------ %%
%
%  TITLE
%
%% ------------------------------------------------------------------------ %%

\title{A homogeneous database of sunspot areas covering
more than 130 years}
%
% e.g., \title{Terrestrial Ring Current:
% Origin, Formation and Decay $\alpha\beta\Gamma\Delta$}

%% ------------------------------------------------------------------------ %%
%
%  AUTHORS AND AFFILIATIONS - 3 methods
%
%% ------------------------------------------------------------------------ %%

% Method 1 (for all journals, except Reviews of Geophysics, which
% should use method 3):
% For three or fewer author/affiliation blocks, use \author{} and \affil{}

\author{L. A. Balmaceda}
 %\affil{Group of Astronomy and Space Science, University of Valencia, Spain}
% University of Arizona, Tucson, Arizona, USA}

\author{S. K. Solanki}
% \affil{Department of Geography, Ohio State University,
% Columbus, Ohio, USA}

\author{N. A. Krivova}
\affil{Max-Planck-Institut f\"{u}r Sonnensystemforschung -- Max-Planck-Str. 2, 
37191 Katlenburg-Lindau,
Germany}

\author{S. Foster}
\affil{Space And Atmospheric Physics Group, Blackett Laboratory, Imperial 
College London, Prince Consort
Road, SW7 2BZ United Kingdom}

\begin{abstract}

The historical record of sunspot areas is a valuable and widely used proxy of 
solar activity and
variability. The Royal Greenwich Observatory (RGO) regularly measured this and 
other parameters between
1874 and 1976. After that time records from a number of different observatories 
are available. These,
however, show systematic differences and often have significants gaps. Our goal 
is to obtain a uniform
and complete sunspot area time series by combining different data sets. A 
homogeneus composite of
sunspot areas is essential for different applications in solar physics, among 
others for irradiance
reconstructions. Data recorded simultaneously at different observatories are 
statistically compared in
order to determine the intercalibration factors. Using these data we compile a 
complete and
cross-calibrated time series. The Greenwich data set is used as a basis until 
1976, the Russian data (a
compilation of observations made at stations in the former USSR) between 1977 
and 1985 and data
compiled by the USAF network since 1986. Other data sets (Rome, Yunnan, Catania) 
are used to fill up
the remaining gaps. Using the final sunspot areas record the Photometric Sunspot 
Index is calculated.
We also show that the use of uncalibrated sunspot areas data sets can seriously 
affect the estimate of
irradiance variations. Our analysis implies that there is no basis for 
the claim that UV irradiance variations
have a much smaller influence on climate than total solar irradiance variations.

\end{abstract}

%% ------------------------------------------------------------------------ %%
%
%  BEGIN ARTICLE
%
%% ------------------------------------------------------------------------ %%

% The body of the article must start with a \begin{article} command,
% and an \end{article} command must be placed at the end of the file,
% before \end{document}.
%
% If using draft mode \end{article} must follow the references section.

\begin{article}

%% ------------------------------------------------------------------------ %%
%
%  TEXT
%
%% ------------------------------------------------------------------------ %%

\section{Introduction}

\label{intro}

The total area of all sunspots visible on the solar hemisphere is one of the 
fundamental indicators of
solar magnetic activity. Measured since 1874, it provides a proxy of solar 
activity over more than 130
years that is regularly used, e.g., to study the solar cycle or to reconstruct 
total and spectral
irradiance at earlier times 
\citep[e.g.,][]{brandt94,solanki98b,li99,li05,preminger05, krivova07}. 
Consequently, a
reliable and complete time series of sunspot areas is essential. Since no single 
observatory made such
records over this whole interval of time, different data sets must be combined 
after an appropriate
intercalibration. In this sense, several comparative studies have been carried 
out in order to get an
appropriately cross-calibrated sunspot area data set \citep[see, for 
instance:][and references
therein]{hoyt83, sivaraman93, Fligge97, Baranyi01, foster04}. They pointed out
that differences between data
sets can arise due to random errors introduced by the personal bias of the 
observer, limited seeing
conditions at the observation site, different amounts of scattered light, or the 
difference in the time
when the observations were made. Systematic errors also account for a disparity 
in the area
measurements. They are related to the observing and measurement techniques and 
different data reduction
methods. For example, areas measured from sunspot drawings are on
average smaller than the ones measured from photographic plates 
\citep{Baranyi01}.

In this work, we compare data from Russian stations and the USAF (US Air
Force) network as well as from other sources (Rome, Yunnan, Catania) with Royal 
Greenwich Observatory (RGO)
data. This combination provides a good set of observations almost free of gaps 
after 1976. Combining
them appropriately improves the sunspot area time series available at present. 

In Section~\ref{data} we
describe the data provided by the different observatories analyzed here. The 
method to calculate the
apropriate cross-calibration factors is explained in Section~\ref{analysis}. The 
results of the
different comparisons are presented in Section~\ref{results}. In 
Section~\ref{psi} we discuss one
central application of sunspot areas: solar irradiance reconstructions. When 
sunspots pass across the
solar disc, a noticeable decrease in the measured total solar irradiance is 
observed.  This
effect can be quantified by the photometric sunspot index \citep{willson80,foukal81,hudson82}. 
This index depends on the positions of the sunspots on the visible solar 
disc and on the
fraction of the disc covered by the spots, i.e. on the sunspot area. It is thus 
clear that
appropriately cross-calibrated sunspot areas are required for accurate 
reconstructions of solar
irradiance \citep[see][]{froehlich94,Fligge97}. We compare results when raw and 
calibrated data are
used to calculate this index and total solar irradiance in Section~\ref{foukal}. 
Finally, Section
\ref{summary} presents the summary and conclusions.

%__________________________________________________________________
\section{Observational data}\label{data}
Data from RGO provide the longest and most complete record of sunspot areas.
The data were recorded at a small network of observatories (Cape of Good
Hope, Kodaikanal and Mauritius) between 1874 and 1976, thus covering nine
solar cycles. Heliographic positions and distance from the central meridian
of sunspot groups are also available.\

The second data set is completely independent and was published by the 
\textit{Solnechniye Danniye}  (Solar Data) Bulletin issued by the Pulkovo 
Astronomical Observatory.
The data were obtained at stations belonging to the
former USSR. These stations provided sunspot areas corrected for
foreshortening, together with the heliographic position (latitude,
longitude) and distance from center of solar disc in disc radii for each sunspot 
group. We will refer to this data set as Russian data.

After RGO ceased its programme, the US Air Force (USAF) started compiling
data from its own Solar Optical Observing Network (SOON). This network
consists of solar observatories located in such a way that 24-hour synoptic
solar monitoring can be maintained. The basic observatories are Boulder and
the members of the network of the US Air Force (Holloman, Learmonth, Palehua,
Ramey and San Vito). Also, data from Mt. Wilson Observatory are included. This
programme has continued through to the present with the help of the US National
Oceanic and Atmospheric Administration (NOAA). This data set is referred to by
different names in the literature, e.g., SOON, USAF, USAF/NOAA, USAF/Mt.
Wilson. In the following, we will refer to it as SOON.\

Usually multiple measurements are provided for a given sunspot group \textit{n} on a particular day \textit{d} coming from different SOON stations, up to a maximum of 6 if all the stations provided information. Normally, at least three values are listed.  In order to get a unique value for this group, we calculate
averages of sunspot areas recorded on this day \textit{d} ($A_{n,d}$),
including only those values which fulfill the following condition:\\
$\overline{A_{n,d}}-2\cdot\sigma_{A}\leq A_{n,d}\leq 
\overline{A_{n,d}}+2\cdot\sigma_{A}$. \\
Here, $\overline{A_{n,d}}$ is the mean value of all the areas measured for
the group \textit{n} on the day \textit{d}, and $\sigma_{A}$ is their standard
deviation. The value of $\sigma_{A}$ varies from group to group, depending on their sizes and 
on the time of the solar cycle. Note that, by using this condition we intend to 
exclude outliers, i.e. those areas whose values differ greatly from the mean for 
each group. In those cases where the number of stations providing data is 1, the 
area for that group is taken from this single source. If the number is 2, the 
area for that group is the mean of the areas measured by both stations.

After that, sunspot areas for
individual groups are summed up to get the daily value. Also averaged are
latitudes and longitudes of each sunspot group recorded by those
observatories whose data are employed to get the mean sunspot area.

These three data sets (RGO, Russia and SOON) are the prime sources
of data that we consider, since they are the most complete, being
based on observations provided by multiple stations. A number of
further observatories have also regularly measured sunspot areas
during the past decades. The record from Rome Astronomical Observatory, whose
measurements began in 1958, covers more than three consecutive and
complete solar cycles. It has several years of observations in
common with Russian stations and SOON as well as with RGO. This is perhaps the
only source of data with a long period of overlap with all three
prime data sets. The database from Rome is used to compare the
results obtained from the other observatories and also to fill up
gaps whenever possible. Unfortunately, its coverage is
limited by weather conditions and instrumentation problems. Whenever
available, data from Yunnan Observatory in China and Catania
Astrophysical Observatory in Italy are also used to fill up the
remaining gaps. In Catania, daily drawings of sunspot groups were
made at the Cooke refractor on a 24.5 cm diameter projected image
from the Sun, while the measurements provided by the Chinese
observatory are based on good quality white light photographs. Table
\ref{Table1} summarizes the information provided by each
observatory: the period in which observations were carried out, the
observing technique, the coverage (i.e. the percentage of days on
which measurements were made) and the minimum area reported by each
observatory. Areas corrected for foreshortening are provided in all
the cases. Directly observed or projected areas, can be
derived using the heliographic positions for sunspot groups and
hence heliocentric angle, $\theta$, or $\mu$-values ($\cos\theta$). Striking
is the relatively large minimum area considered by the SOON network. This
suggests that many smaller sunspots are neglected in this record.

All the data used in this work were extracted from the following website: 
\newline
http://www.ngdc.noaa.gov/stp/SOLAR/ftpsunspotregions.html. 

%__________________________________________________________________

\section{Analysis}

\subsection{Cross-calibration factors}\label{analysis}

Daily sunspot areas from two different observatories are directly compared on
each day on which both had recorded data. We deduced multiplication factors
needed to bring all data sets to a common scale, namely that of RGO, which is
employed as fiducial data set.

For this, the spot areas from one data set are plotted vs the other (see left 
panels of
Fig.~\ref{fig1}). The slope of a linear regression forced to pass through the 
origin (see Appendix~\ref{offset}) 
can be used to calibrate the sunspot area record considered auxiliary, 
$A^{aux}$, to the
areas of another basic data set, $A^{bas}$:
\begin{equation}
A^{bas}=b\cdot A^{aux}.
\end{equation}

First, this analysis is applied to all the points. The slope thus obtained is
taken to be the initial estimate for a second analysis where not all the
points are taken into account. Outliers are
excluded by taking only points within 3$\sigma_{fit}$ from the first fit, where
$\sigma_{fit}~=~\sqrt{\frac{1}{N-1}\sum_{i=1}^{N}\left( A_{i}^{bas}-b\cdot
A_{i}^{aux}\right)^2 }$.
Also, only areas lying above the line joining the
points (0, $3\sigma_{fit}$) and ($3\sigma_{fit}$, 0) are considered. Through this
measure points close to
the origin are excluded since they introduce a bias.

Ordinary least-square regression cannot be applied in this case, for the following
reasons: (1) the distinction between independent and dependent variables is arbitrary; 
(2) the data do not provide formal errors for the measurements; (3) the intrinsic
scatter of the data may dominate any errors arising from the measurement procedure of
sunspot areas. A method that treats the variables symmetrically should be used instead.

To this purpose, the same procedure is repeated after interchanging the data sets taken as a
basis and as auxiliary. For the reasons outlined in Appendix~\ref{offset},
the inverse value of the slope now obtained, $b^{\prime}$, differs from
the slope $b$ obtained in the first place. Therefore, the final calibration
factor is then calculated by averaging these two
values: \textit{b} and 1/$b^{\prime}$. This method is referred to as "bisector line" 
\citep{isobe90}.

An alternative method to find the calibration factors is described
in Appendix~\ref{otromet}. This second method does not neglect the
sunspot areas close to zero. In contrast, it gives equal weight to
all values. The calibration factors obtained in this way are thus
less accurate during high activity levels, when solar irradiance is
most variable. Since the reconstruction of solar irradiance is a key
application of the new cross-calibrated sunspot area record, we
select the method described above rather than the one presented in
Appendix~\ref{otromet}. Of course, for other applications, this method may
happen to be more appropriate. Therefore, in Table~\ref{Table5} we
also give the factors obtained in this way. The difference between the factors obtained by the 2 methods is generally less than 5\%, although differences as large as 12\% can be reached for factors deduced from corrected sunspot areas.

Data series that do not overlap in time can be intercalibrated using the Zurich 
sunspot number as a
common index \citep{Fligge97, vaquero04}. Since this approach requires an 
additional assumption, namely
that the size distribution of sunspots \citep{bogdan88,baumann05} remains 
unchanged over time we avoid
using it for calibration purposes \citep{solanki04c}. We use this comparison 
only for confirmation of
the results obtained from the direct measurements, so that the new record is
completely independent of
the sunspot number time series.

\subsection{Error estimates}\label{error}

A single calibration factor is calculated for the whole period of overlap
between data sets obtained by two observatories. This is repeated once for
the projected areas and those corrected for foreshortening provided by the different observatories.

\begin{figure*}%[p!]%[t!]
\resizebox{\hsize}{!}{\includegraphics{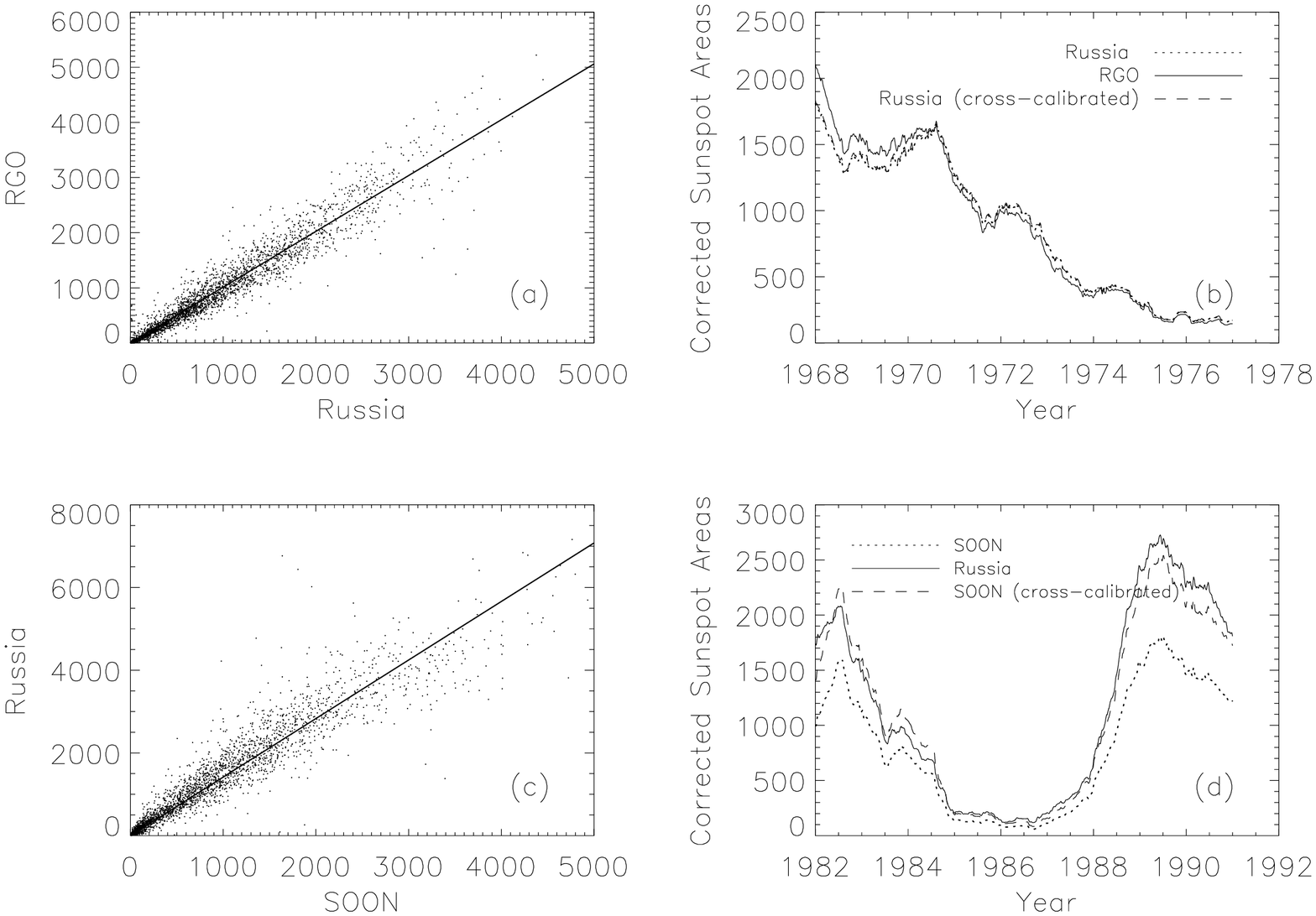}}
\caption{\footnotesize Comparison of sunspot areas corrected for foreshortening 
obtained by different
observatories. Top: RGO vs Russia, bottom: Russia vs SOON. Left: scatter plots. 
Solid lines represent
linear regressions to the data neglecting a possible offset (i.e., forced to 
pass through zero) as well
as data points close to the origin and the outliers lying outside the $\pm 
3\sigma$ interval from the
fit. Right: 12-month running means of sunspot areas vs time. Solid curves show 
the data used as basis
level, dotted are the data from the second observatory and dot-dashed the 
calibrated areas.}
\label{fig1}
\end{figure*}

 In some cases, however, the relation between two data sets was found to
evolve with time. This can be seen in the right panels of Fig.~\ref{fig1}, in 
which the 12-month running
means of sunspot area records are plotted vs time. Both Figs.~\ref{fig1} and 
\ref{fig2} show that even
after cross-calibration the two data sets do not run in parallel but rather have 
systematic relative
offsets over particular periods of time (lasting multiple years). Therefore 
factors for different
sub-intervals are also calculated, in order to estimate the uncertainties of the 
final factors. This is
performed by separating different solar cycles. When the whole interval of 
overlap does not cover more
than one cycle, then the division is made when a change in the behaviour is 
observed. See, e.g., the
comparison between RGO and Russian data in the upper right panel of 
Fig.~\ref{fig1}, where the change
takes place after year 1971. Before that year, Russian areas are on average 
smaller than those from
RGO, while the situation is reversed afterwards. The uncertainty in the final 
factors is thus the combination of the uncertainties due to the cycle-to-cycle 
variations (different factors for different cycles and/or subintervals), 
$\sigma_{cyc}$,
the difference between $b$ and $b^{\prime}$, $\sigma_{dif}$, and the errors, 
$\sigma_{slope}$, in determining the
slopes so that:
$\sigma=\sqrt{{\sigma_{cyc}}^2+{\sigma_{dif}}^2+{\sigma_{slope}}^2}$.
 The main source of uncertainties being the fact that the relationship between 
two
given observatories during the period they overlap is not uniform. Therefore, 
the smallest errors are
obtained when this period is short (see, e.g., Russia -- Catania, SOON -- 
Catania in Table~\ref{Table4}).
On the other hand, the largest errors are found in the comparison between Russia 
and Yunnan. This is
discussed in more detail in the next Section.

%
%______________________________________________________________

\section{Results and discussion}\label{results}

\subsection{Comparison between sunspot areas}\label{comparison}

The results of the analysis described in Section 3 are summarized in Table~\ref{Table4}. The first two columns give the names of the data sets being
compared. The observatories whose data are taken as the basis are indicated as
Obs.~1, while the observatories whose data are recalibrated are indicated as
Obs.~2. The third column shows the interval of time over which they overlap. In 
the
next two columns we list the calibration factors by which the data of
Obs.~2 have to be multiplied in order to match those of Obs.~1. The
factors for the originally measured areas (projected areas, PA) and 
for the areas corrected for foreshortening (CA) are given, in columns 5 and 6, 
respectively.
The two last columns list the corresponding correlation coefficients
between the two data sets.

With one exception the correlation coefficients for the projected areas are
larger than for the ones corrected for foreshortening. This is not
unexpected, since errors in the measured position of a sunspot increase the
scatter in the areas corrected for foreshortening, while leaving the projected
areas unaffected.

In the following we discuss the results in greater detail. 
The overlap between RGO and Russian data covers the descending phase of cycle~21.
As can be seen from Figs.~\ref{fig1}a and b the two sets agree rather
well with each other. The cross-calibration factor is very
close to unity, although the difference between the two data sets displays a 
trend with time. Before
1971, areas from RGO are larger (6\% for projected, 8\% in case of corrected for 
foreshortening) than
Russian measurements, whereas after that time
areas from the Russian data set are 8\% larger (see Fig.~\ref{fig1}b) for both, 
projected and corrected areas. This trend
remains also after recalibrating the Russian data, because a single factor is 
not sufficient to remove
this effect. Since it is not clear which (or both) of these two data sets 
contains an artificial drift,
we do not try to correct for it.

Russian and SOON areas display more significant differences (see Fig.~\ref{fig1} 
c and d). The overlap
covers the period from 1982 to 1991, or cycles 21 and 22. During the whole time 
interval, SOON areas
appear to be smaller (by on average 40\% for projected and 45\% for the corrected ones) than those
of the Russian data. This is mainly due to the
significant difference in the minimum value of the counted sunspots (1 ppm of 
the solar hemisphere for
Russian vs 10 ppm of the solar hemisphere for SOON observations, see 
Table~\ref{Table1}). As can be seen from
Fig.~\ref{fig1}d, data from these two records also do not run in parallel, 
exhibiting quite a
significant trend relative to each other (compare solid and dashed curves).

In general, it was found that areas measured by the SOON network as well as
those by the Rome, Catania and Yunnan observatories are on average
smaller than areas reported by RGO and Russian stations, which agrees with
the fact that the minimum areas of individual spots included into these two
records are the smallest. For the same reason, SOON areas are smaller on
average than the measurements from other data sets: the minimum area of the recorded spots is a factor of 3 to 10 higher for SOON than for the other observatories.

The last three lines of Table~\ref{Table4} give the factors by which SOON, Catania and Yunnan data 
need to be multiplied
in order to match the RGO data. Since none of these data sets overlap with RGO 
we have used the Russian
data as intermediary. Of course, correlation coefficients can not be determined 
in this case.
The factor needed to calibrate SOON data to the RGO data set is
1.43 for projected areas, in good agreement with the results by 
\citet{hathaway02} and
\citet{foster04}, who both give 1.4. In the case of areas corrected for 
foreshortening the factor found here is $\sim$7\% larger, being 1.49.

\begin{figure}%[p!]%[t!]
\resizebox{\hsize}{!}{\includegraphics{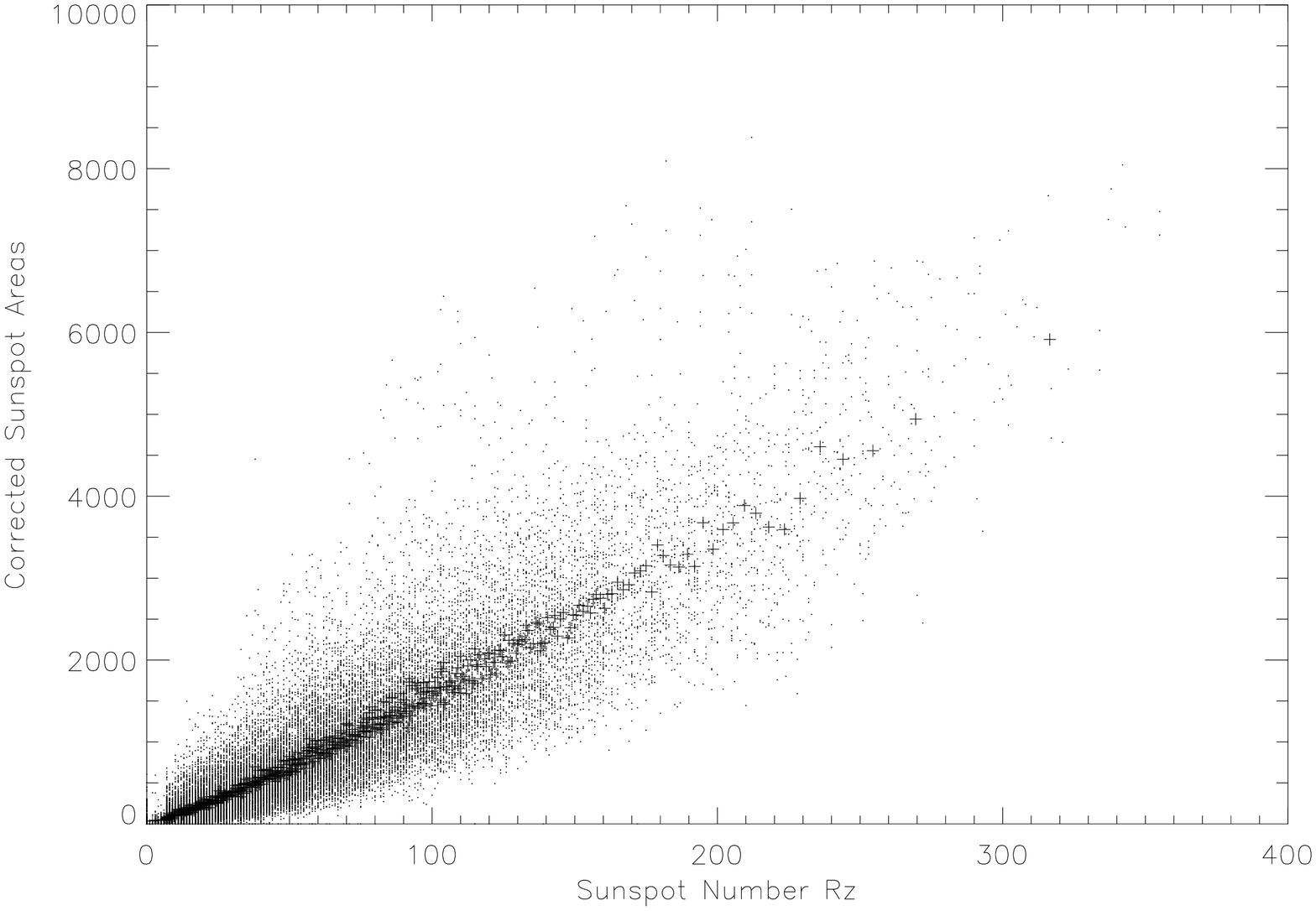}} 
\caption{\footnotesize Sunspot areas
corrected for foreshortening vs sunspot number, $R_z$, for measurements made by 
RGO. The dots represent
daily values between 1874 and 1976. Each '+' symbol represents an average over 
bins of 50 points. }
\label{fig2}
\end{figure}

\subsection{Comparison with sunspot number}\label{sunspotnum}

The relationship between the Zurich relative sunspot number, $R_z$, and sunspot 
area (from a single
record) shows a roughly linear trend with a large scatter. In Fig.~\ref{fig2} we 
plot sunspot areas
corrected for foreshortening, $A_S$, for RGO measurements vs $R_z$. We have 
chosen $A_S$ from RGO since
this is the longest running data set. The plus signs represent data points 
binned in groups
of 50. These points indicate that the relationship is roughly, but not exactly 
linear. In particular at
low $R_z$ values, $A_S$ appears to be too small, possibly because of the cutoff 
in the $A_S$
measurements. However, this behaviour may reflect also the particular definition
of $R_z=k(10g+s)$, where \textit{g} is the number of sunspot groups, \textit{s} the total
number of distinct spots and \textit{k} the scaling factor (usually $ <~1$) which
depends on the observer and is introduced in order to keep the original
scale by Wolf \citep{waldmeier}. In this definition, even a
small group of sunspots is given a nearly equally large weight as a large group.
It is observed from
this plot that a given value of $R_z$ corresponds to a range of values of 
sunspot areas. However, the
scatter due to points within a single cycle is larger than the scatter from 
cycle to cycle.

When studying the relationship between $A_S$ and $R_z$ for individual cycles, it 
was observed that in
some cases the scatter is significantly higher. In such cycles, large areas are 
observed while $R_z$
remains low. In particular, the shape of $A_S$ cycles resembles that of $R_z$ 
cycles, but individual
peaks are more accentuated in $A_S$. This could be also a consequence of the 
definition of $R_z$,
regarding the large weight given to the groups. \citet{Fligge97} already showed 
that, in general, the
relationship between $A_S$ and $R_z$ changes only slightly from one cycle to the
next, with the
difference being around 10\%.

\begin{figure}%%[p!]%[t!]
\resizebox{\hsize}{!}{\includegraphics{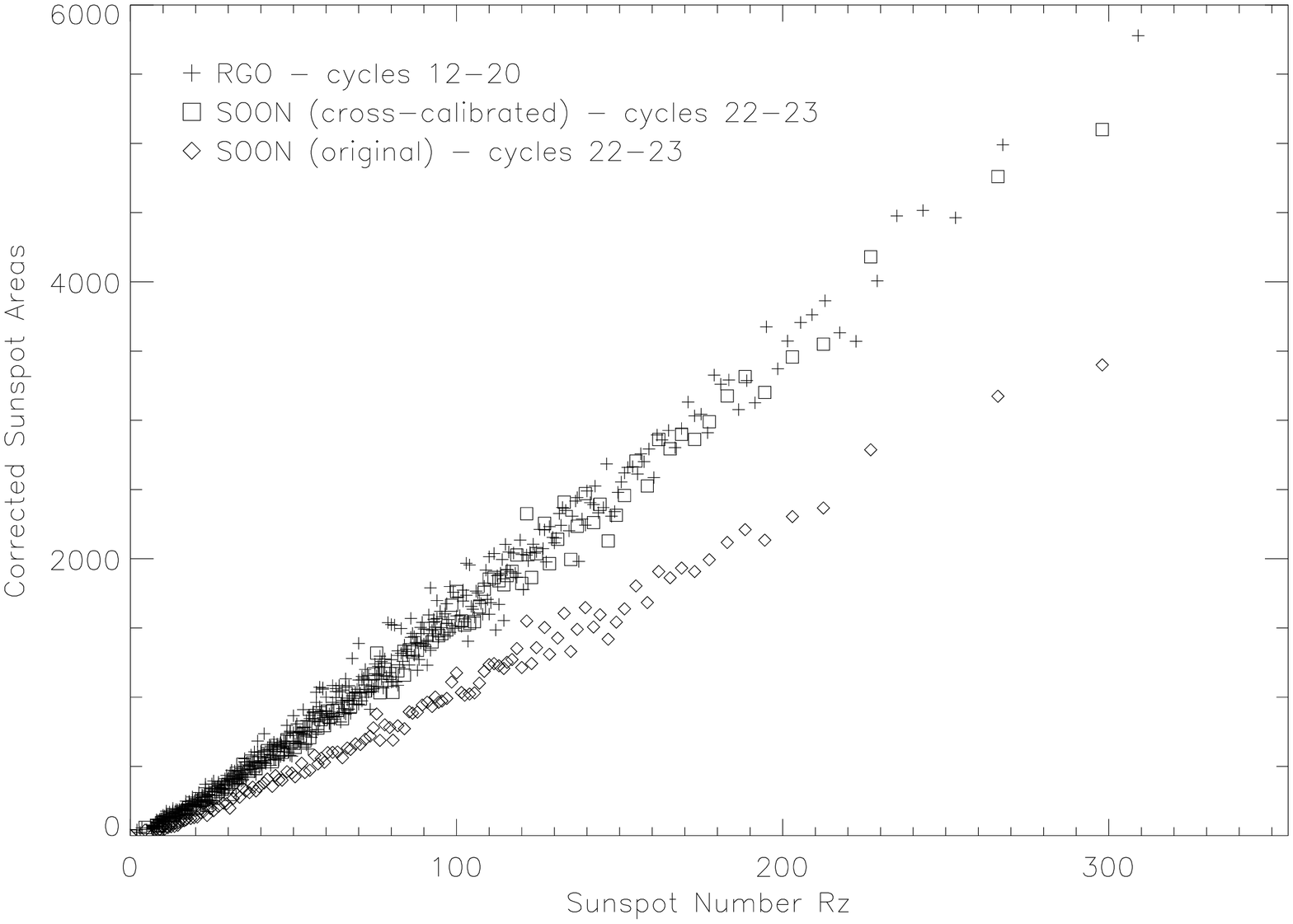}} 
\caption{\footnotesize Sunspot areas
corrected for foreshortening vs Zurich relative sunspot number, $R_{z}$, for 
measurements made by RGO
and SOON. Each symbol represents an average over bins of 50 points. The original 
data from SOON are
represented by diamonds, and those after multiplication by a calibration factor 
of $\sim$1.5 in order to
match RGO data, by squares. } \label{fig3}
\end{figure}

Fig.~\ref{fig3} shows the comparison between RGO and SOON sunspot areas 
corrected for foreshortening with the sunspot number. We have binned the data 
from each observatory every 50 points
according to the sunspot number. The uncalibrated areas from SOON lie significantly below the ones from RGO. After multiplying SOON data by the calibration factor of 1.49 found in Sect.~\ref{comparison}, they display practically the same relationship to $R_{\rm z}$ as the RGO data.

\subsection{Cross-calibrated sunspot area records}\label{cross}

In a next step we create records of projected and corrected sunspot areas 
covering the period from 1874
to 2008 that are consistently cross-calibrated to the RGO values. We use RGO, 
Russia and SOON
measurements as the primary sources of data. As shown by Table~\ref{Table1}, 
these sources provide the
sets of sunspot area measurements, with the least number of gaps.

The individual periods of time over which each of these is taken as the
primary source are:

\begin{tabular}[l]{ll}
& \\
1874 -- 1976 &  RGO \\
& \\
1977 -- 1986 &  Russia \\
& \\
1987 -- 2008 &  SOON\\
& \\
\end{tabular}

\begin{figure}%[p!]%[t!]
\begin{center}
\resizebox{\hsize}{!}{\includegraphics{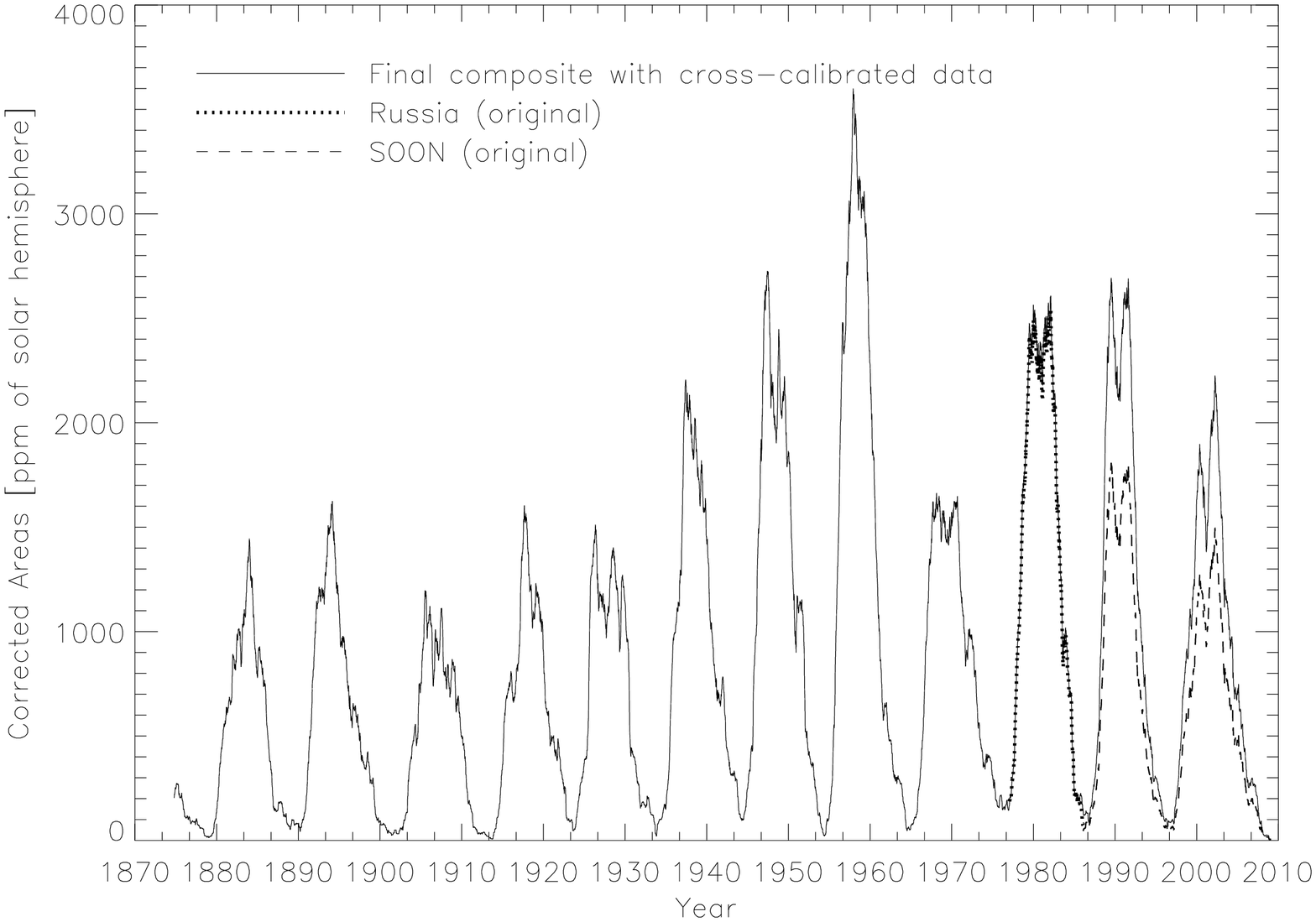}}
\caption{\footnotesize 12-month running means of sunspot areas corrected for
foreshortening of the final composite using the factors given in Table
\ref{Table4} (solid curve). Also plotted are the Russian and SOON data
entering the composite prior to calibration.} \label{fig4}
\end{center}
\end{figure}

The final sunspot area composite is plotted in Fig.~\ref{fig4} (solid curve), and is tabulated 
in Table 4 (only
available electronically). We have chosen to use the Russian data set until 1986 
for the simple reason
that this year corresponds to the solar minimum. In this way, each data set 
describes different solar
cycles (see Fig.~\ref{fig4}). We are aware that this is only aproximately correct since sunspots 
from consecutive cycles
overlap during a short period of time, but this is a second order effect. In 
this combination we opt to
multiply the post-RGO measurements by the factors obtained here since RGO areas 
data set is by far the
longest running and relatively homogeneous source. Any data gaps in the primary 
source are filled using
data from one of the other two primary records (if available), or data from Rome and Yunnan, properly recalibrated. The two last-named series allowed us to fill up the gaps over a total of 115 days. In this way, gaps in the final composite cover only $\sim8$\% of the total length of the combined data set of 49308 days.

\section{The Photometric Sunspot Index}\label{psi}

The passage of sunspots across the
solar disc causes a decrease in the total solar irradiance. This
effect can be quantified by estimating the photometric sunspot
index, $P_S$,  \citep{hudson82}. First, the deficit of radiative flux, $\Delta S_S$, due to
the presence of a sunspot of area $A_S$ is calculated as:
\begin{equation}\label{eq1}
  \frac{\Delta S_S}{S_Q} = \frac{\mu A_S(C_S-1)(3\mu +2)}{2}.
\end{equation}
This value is expressed in units of $S_Q$, the solar irradiance for the quiet 
Sun (i.e. solar surface
free of magnetic fields). $S_Q = 1365.5 \rm{W/m^2}$ is taken from the PMOD 
composite of measured solar
irradiance \citep{froehlich03, froehlich06}. 
We use 
the areas composite obtained here, $A_S$, and
the heliocentric positions, $\mu$, of the sunspots present on the
solar disc. The residual intensity contrast of
the sunspot relative to
that of the background photosphere $C_S - 1$ is taken from \citet{brandt92}. 
It takes into account the dependence of  the sunspot residual intensity contrast on sunspot area,
i.e., larger sunspots are darker
than smaller spots, as has recently been confirmed on the basis of MDI
data by \citet{mathew07}.
Following \citet{brandt92,brandt94} and \citet{froehlich94}  we use:
\begin{equation}\label{eq2}
  C_S - 1 = 0.2231 + 0.0244 \cdot \log(A_S).
\end{equation}

\begin{figure}%[p!]%[h!]%[t!]
\resizebox{\hsize}{!}{\includegraphics{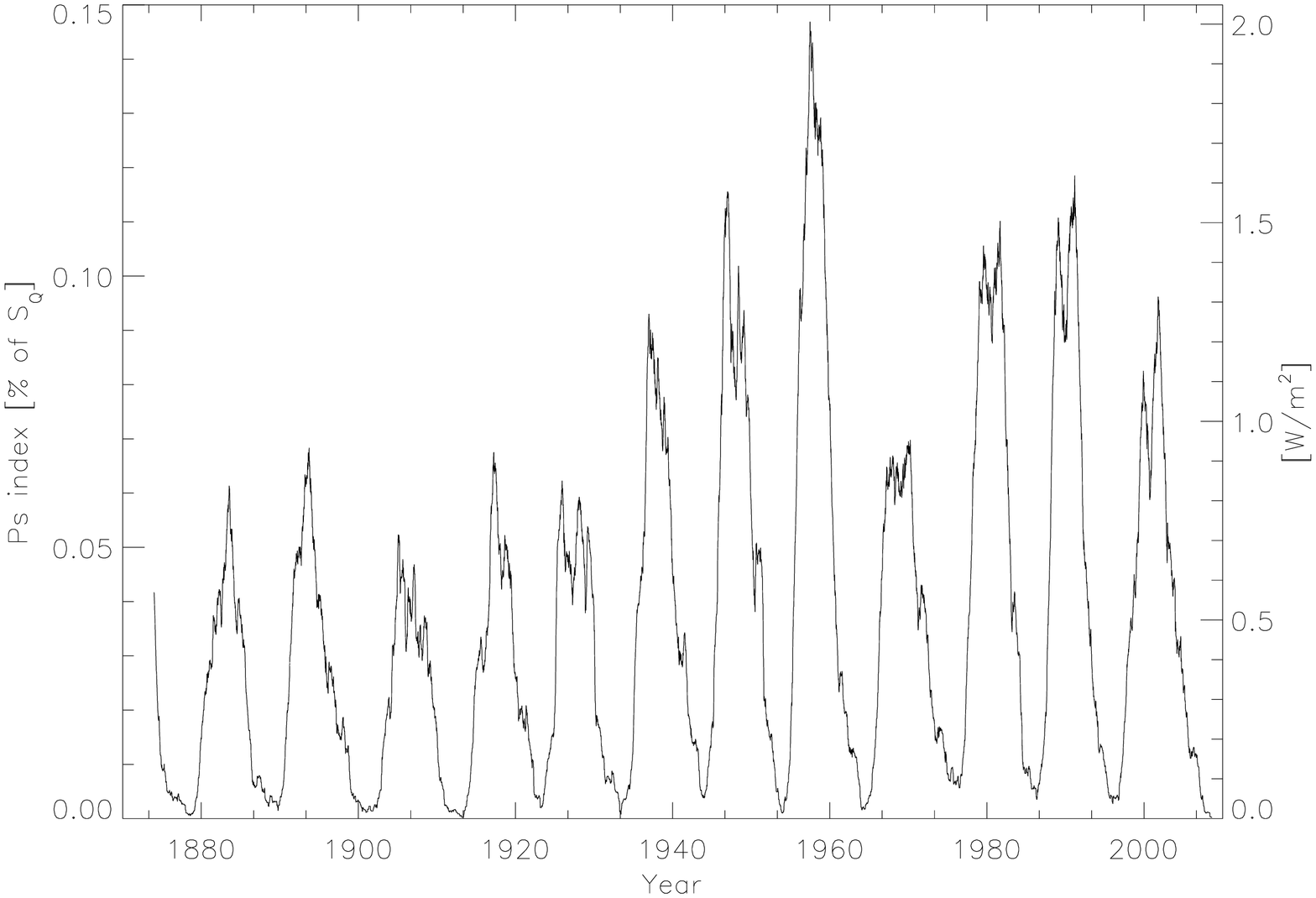}} 
\caption{\footnotesize 12-month running mean
of the photometric sunspot index, $P_S$, computed using the sunspot areas 
composite produced here. The
y-axis is expressed in percents of $S_Q$, the irradiance of the quiet Sun. } 
\label{fig5}
\end{figure}

Finally,  summing the 
effects from all the
sunspots present on the disc we obtain:
\begin{equation}\label{eq3}
 P_S  =\sum_{i=1}^{n} \left( \frac{\Delta S_s}{S_Q}\right)_i.
\end{equation}
Figure \ref{fig5} shows the 12-month running mean time series of  the $P_S$ 
index for the period
1874 -- 2008. The daily $P_S$ values are also listed in Table 4 (available 
electronically).

\section{An example of errors introduced by an uncritical use of uncalibrated 
sunspot areas data sets}\label{foukal}

Variations of solar irradiance on time scales longer than approximately a day 
are caused by the passage
of dark sunspots and bright faculae across the solar disc. Due to the different 
wavelength dependences
of their contrasts, the contribution of faculae is higher in the UV than in the 
visible or IR, whereas
the contribution of sunspots dominates increasingly with increasing wavelength
\citep{solanki98a,unruh99}. Thus employment of a faulty or inconsistent sunspot or 
faculae time series to
reconstruct solar total and UV irradiance can lead to systematic differences 
between them.

Now, it has been claimed that variations of solar UV irradiance are less 
important for climate than
variations of solar total irradiance, \textit{S}, \citep{foukal02,foukal06}. 
These results are based on
uncalibrated sunspot areas including both the Greenwich and the SOON data sets. 
Here we show that when
sunspot areas after appropriate intercalibration as desribed in 
Section~\ref{analysis} are employed,
total and UV solar irradiance behave similarly.

We redo the analysis of \citet{foukal02},
but employing the cross-calibrated time series of sunspot areas obtained
here. For the facular contribution we employ the same proxy as
\citet{foukal02}, a monthly mean time series of plage plus enhanced
network areas, $A_{PN}$. This data set was kindly provided by P.
Foukal. Areas were measured from
spectroheliograms and photoheliograms in the K-line of Ca II
obtained at Mt. Wilson, McMath-Hulbert and Big Bear observatories in
the period 1915 -- 1984 \citep{foukal96,foukal98}. Later, this time
series was extended until 1999 using data from Sacramento Peak
Observatory (SPO). The data cover the period August 1915 --
December 1999 inclusive. The identification of plages and enhanced
network was performed by several observers. Details about the reduction procedure to derive the
$A_{PN}$ index can be found in \citet{foukal96}. $A_{PN}$ values are expressed
in fractions of
the solar disc. 

Total and UV solar irradiance time series are reconstructed following 
\citet{foukal02}. According to
that approach, enhancements in total solar irradiance are proportional to the 
difference in plage,
$A_{PN}$, and sunspot areas, $A_S$, whereas enhancements in UV irradiance are 
proportional to the plage
areas alone. As a first step, residuals of solar irradiance after removing the 
sunspot darkening, $S -
P_S$, are calculated for the time when irradiance measurements are available, 
i.e. from 1978 till
present. This quantity, $S - P_S$, is a measure of facular contribution to the 
total irradiance. Total
solar irradiance measurements, $S$, are taken from the PMOD composite derived 
from different
instruments with best allowance for their degradation and inter-calibration 
\citep{froehlich00,
froehlich06}. Then, a regression relation of the form: $S - P_S = b \cdot A_{PN} 
+ a$ is constructed
between the monthly mean values of these residuals and of the plage areas, 
$A_{PN}$. This regression
relation is then used to reconstruct the residuals $(S - P_S)_{rec}$ between 
1915 and 1999 when values
of $A_{PN}$ are available. The reconstructed total solar irradiance is finally 
obtained by just adding
back the time series of $P_S$ over this period.

\begin{figure}%[p!]%[h!]%[t!]
\resizebox{\hsize}{!}{\includegraphics[angle=90]{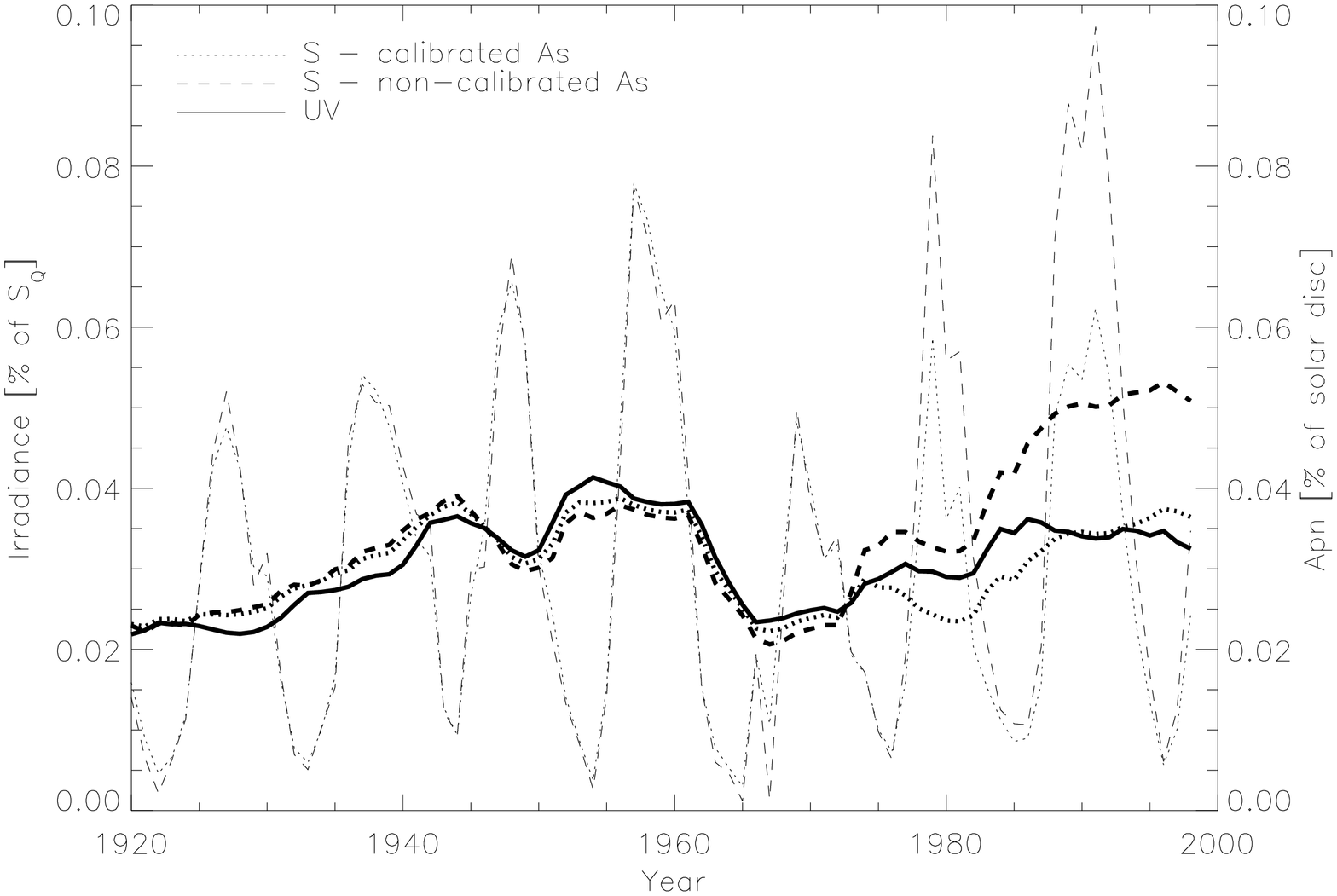}} 
\caption{\footnotesize 11-yr running
mean of total solar irradiance based on calibrated areas (dotted line), 
non-calibrated areas (dashed line) and UV irradiance given by the variations in the $A_{PN}$ (solid 
line). The thin lines represent 1-year means of solar irradiance based on calibrated areas (dotted line) and non-calibrated areas (dashed line).
The y-axis is
expressed in percents of $S_Q$, the irradiance of the quiet Sun. } \label{fig6a}
\end{figure}

Figure \ref{fig6a} shows the 11-yr running means of the reconstructed total 
irradiance using calibrated
(thick dotted line) and non-calibrated (thick dashed line) data. The thin lines represent the 1-yr means of both reconstructions.
The curves were scaled 
in order to highlight
the difference in the upward trend after 1970. The dashed curve represents the 
UV irradiance, i.e., the
solar flux at wavelengths shorter than 250 nm. Its variability
 is determined mainly by the bright magnetic plages in active regions and
enhanced network produced as these regions decay. Its reconstruction follows
the same steps as of the total solar irradiance, except that the last step
(adding back the $P_S$) is not carried out.

The total irradiance reconstructed by \citet{foukal02}, which is very similar to 
the grey curve in
Fig.~\ref{fig6a}, shows a clear upward trend after the year 1976 due to the 
strong presence of faculae
that is not balanced by increased sunspot area. 
The UV irradiance does not 
display such a prominent
rise, however. This result was interpreted by \citet{foukal02} as evidence 
for a strongly different
behaviour of the total irradiance and UV irradiance and consequently their very 
different influence on the
Earth's climate. In particular, the fact that the TSI correlates much better 
with global climate than
the UV irradiance during the last three decades led \citet{foukal02} to propose 
that UV irradiance
influences global climate less than total irradiance. However, we find here that 
this behaviour is no
longer observed when appropriately calibrated areas are used. The shape of the 
total irradiance
estimated from calibrated data now follows closely the shape of the variation in 
$A_{PN}$, i.e. the UV
irradiance \citep[cf.][]{solanki03}. It is not by chance that the two reconstructions of $S$ start to 
diverge in $\sim$1976
since at that time the record of $A_{S}$ from RGO ends.

We stress that the simple approach used here to reconstruct total and UV solar 
irradiance has
shortcomings. One concerns the $A_{PN}$ time series, which is based on 
uncalibrated spectroheliograms.
Film calibration in photographic plates and variable image quality are some of 
the factors that
introduce uncertainties in the extraction of the features and need to be taken 
into account. They
affect the correct identification of different features in the CaII K images 
which is based on criteria
of decreasing intensity, decreasing size or decreasing filling factors 
\citep{worden98}. Another
concerns the simplicity of the model assumed here, which succesfully reproduces 
the cyclic variation
but does not contain a secular trend, unlike more detailed and complete recently 
developed models, for
instance: \citet{foster04,wang05,krivova07}. Such a secular trend can be 
produced by long-term changes
in the network, which is only poorly sampled by the $A_{PN}$ data employed here. 
These shortcomings
have no influence on the drawn conclusions, however. It is not the aim of this 
section to produce
realistic records of total and UV irradiance, but rather to demonstrate the 
importance of using a
carefully cross-calibrated sunspot areas time series. In particular, our 
conclusion that total solar
irradiance shows no strong upward trend in three decades since 1976 is supported 
by the irradiance
composite of \citet{froehlich00,froehlich06} and the modelling work of 
\citet{wenzler06}.

\section{Summary and conclusions}\label{summary}

In this work, we have compared sunspot areas measured at different 
observatories. We found a good
agreement between sunspot areas measured by Russian stations and RGO, while a 
comparison of sunspot
areas measured by the SOON network with Russian data shows a difference of about 
40\% for projected
areas and 44\% in areas corrected for foreshortening. This is at least partly 
due to the different
minimum areas of sunspots taken into account in these data sets: smallest areas 
included in the RGO and
Russian records are 10 times smaller than those in the SOON series (see 
Table~\ref{Table1}). Histograms
of sunspot areas show that such small sunspots are rather common 
\citep{bogdan88,baumann05}. SOON
sunspot areas are combined with those from RGO and Russia by multiplying them by 
a factor of 1.43 in
the case of projected areas and 1.49 in the case of areas corrected for 
foreshortening. Data from other
observatories are employed to fill up some of the remaining gaps. In this 
manner, a consistent sunspot
area database is produced from 1874 to 2008. 

A properly cross-calibrated sunspot 
areas data set is
central for, e.g., reliable reconstructions of total and spectral solar 
irradiance. In order to
demonstrate this, we have also presented a simple reconstruction of total and UV 
solar irradiance based
on sunspot and plages plus enhanced network areas for the period 1915 -- 1999. 
We showed that the use
of data of different sources directly combined, without a proper 
cross-calibration can lead to
significantly erroneous estimates of the increase of solar irradiance in the 
last decades. This means
in particular that the claim of \citet{foukal02} that UV solar irradiance is far 
less effective in
driving climate change than total solar irradiance has no basis. \\
Data from 
additional observatories,
such as Debrecen Observatory in Hungary \citep{gyori98,gyori00}, will help to 
improve the sunspot areas
record even further. Another interesting possibility not explored here would be the comparison with data from space-borne observations, which are unaffected by seeing. 
SOHO/MDI 
\citep{scherrer95} provides
 continuous data free of atmospheric effects since 1996 till present.
 \citet{gyori05} and \citet{gyori06} have presented a comparison between areas 
measured
 by Debrecen Observatory and MDI for 1996 and 1997. After applying the same 
procedure for determining
 sunspot areas to both data sets, they found that MDI areas are 17~\% larger. 
They attribute this
 difference to the smaller scale of MDI images, with respect to that of Debrecen 
data.
\citet{wenzler05t}, on the other hand, compared umbrae and penumbrae areas derived from continuum images taken at the Kitt Peak Observatory (KP) and MDI. From the analysis of 24 selected days at different levels of solar activity between 1997 and 2001, he obtained almost identical values for locations and areas for both data sets by applying an appropriate threshold. He also compared total daily KP sunspot areas and the composite presented here. The comparison showed that SPM areas are about 4\% lower for the period 1992-2003 (2055 days). This shows that it is possible to combine ground-based and space-based measurements of sunspot areas into a single time series.

\appendix
\section{On the effect of including offset in the calculation of 
cross-calibration factors }\label{offset}

\begin{figure}%[p!]%[h!]%[t!]
\begin{center}
\resizebox{\hsize}{!}{\includegraphics[angle=90]{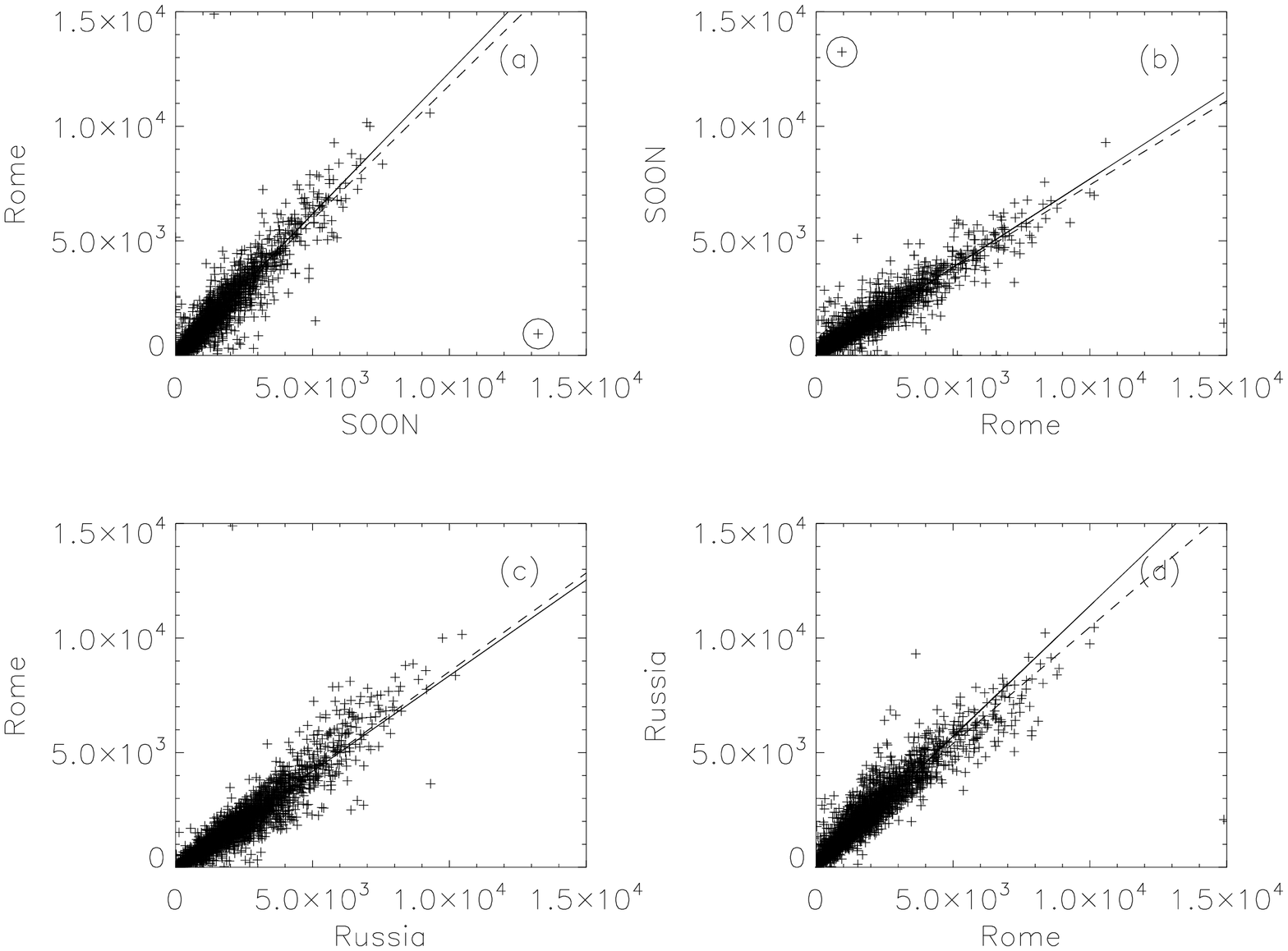}} 
\caption{\footnotesize Comparison between
sunspot areas recorded by different observatories. Areas are in units of 
millionths of the solar disc.
The lines represent linear regressions to the data: standard (dashed) and forced 
to pass through the
origin (solid). The encircled data point is discussed in the text.} \label{fig6}
\end{center}
\end{figure}

Let us consider sunspot areas recorded by two different observatories, Obs.~1 
and Obs.~2, during the same
period. Let \textit{b} be the slope of the linear regression when the area 
recorded by Obs.~2 is the
independent variable and $b^{\prime}$ the slope when the area recorded by Obs.~1 
is the independent
variable. In the ideal case, $b = 1/b^{\prime}$. However for real data sets this 
is not true. There are two
reasons for this. Firstly, since sunspot areas cannot be negative, values close 
to zero introduce a
bias into the regression coefficients. As a result, the slopes we obtain 
including an offset (dashed
lines in Fig.~\ref{fig6}) are typically lower than the ones obtained by 
considering no offset (solid
lines in Fig.~\ref{fig6}). In particular, the obtained \textit{b} is always 
lower than $1/b^{\prime}$, whereas
$b^{\prime}$ is lower than $1/b$. In order to overcome this, we force the fit to 
go through the origin
(solid lines in Fig.~\ref{fig6}). The corresponding slopes typically increase, 
such that values of
\textit{b} and $1/b^{\prime}$ become closer to each other, although they still 
differ. Secondly, when carrying out a
linear regression to the relationship between the observatories, we assume 
measurements by one of them
to be free of errors, whereas in reality both records are subject to errors. 
This immediately produces
different regressions depending on which data set is plotted on the ordinate. 
This is well illustrated
by comparing the encircled data point in Figs. \ref{fig6}a and b (it corresponds 
to the same data point
in both). In Fig.~\ref{fig6}a, the point significantly lowers the regression 
slope, since there are
hardly any data points at that location of the x-axis, while in Fig.~\ref{fig6}b 
its influence is small,
since it now lies at a well populated part of the x-axis. By removing such 
outliers, we further reduce
the difference between $b$ and $1/b^{\prime}$, but they are still not identical 
for purely statistical reasons.
Therefore, as final factors we take the average between \textit{b} and 
1/$b^{\prime}$ \citep{isobe90}.

A more complicated case is the one when there is a significant offset between 
Obs.~1 and Obs.~2, for
example due to the difference in the minimum area of the considered spots (see, 
e.g. Fig.~\ref{fig6}c and
Table~\ref{Table1}). In this case it may happen that \textit{b} is not lower 
than $1/b^{\prime}$. Then the
slopes obtained by forcing the fit to go through zero do not necessarily improve 
the original ones.
However, we apply the same procedure, neglecting the offset, for the following 
reasons: (1) the
magnitude of the offset is rather uncertain due to the bias introduced by the 
positivity of the sunspot
areas; (2) in doing this we may introduce some errors mainly at low values of 
sunspot areas, whereas
values obtained during high activity levels which are of higher priority here 
are on average relatively
reliable; (3) the real slope still lies in the range [$b, 1/b^{\prime}$] (or 
[$b^{\prime}, 1/b$]) so that an average of
\textit{b} and $1/b^{\prime}$ (or $b^{\prime}$ and $1/b$) is a good 
approximation and, finally, (4) there are
only few such cases (like the comparisons between areas from Russia and Rome and 
from Rome and SOON
shown in Fig.~\ref{fig6}).

An additional possible reason for the difference between $b$ and $1/b'$ may be that the true relationship is non-linear. However, the scatter in the data is too large to reach any firm conclusion on this.

\section{An alternative method to calculate cross-calibration 
factors}\label{otromet}
In addition to the method described in Section~\ref{analysis} and 
Appendix~\ref{offset} to
cross-calibrate different sunspot area data sets, we also performed the
cross-calibration by varying a parameter \textit{f} (defined below) in order
to minimize a merit function $\mathcal M$, calculated over the $N$-days on
which both $A^{bas}$ and $A^{aux}$ are available:
\begin{equation}\label{eqb}
\mathcal M=\displaystyle\sum_{i=1}^{N}{\left[A^{bas}_{{t}_{i}}-f \cdot
A^{aux}_{{t}_{i}} \right]^{2}}.
\end{equation}

The merit function is used here since due to the lack of individual errors for daily measurements the classical definition of $\chi^2$ cannot be applied.

In order to find the absolute minimum of $\mathcal M$ irrespective of the 
presence of any secondary
minima, a genetic algorithm called Pikaia is used
\citep[][http://www.hao.ucar.edu/public/research/pikaia/pikaia.html]{charboneau95}.

\begin{figure}%[p!]%[h!]%[t!]
\begin{center}
\resizebox{\hsize}{!}{\includegraphics[angle=90]{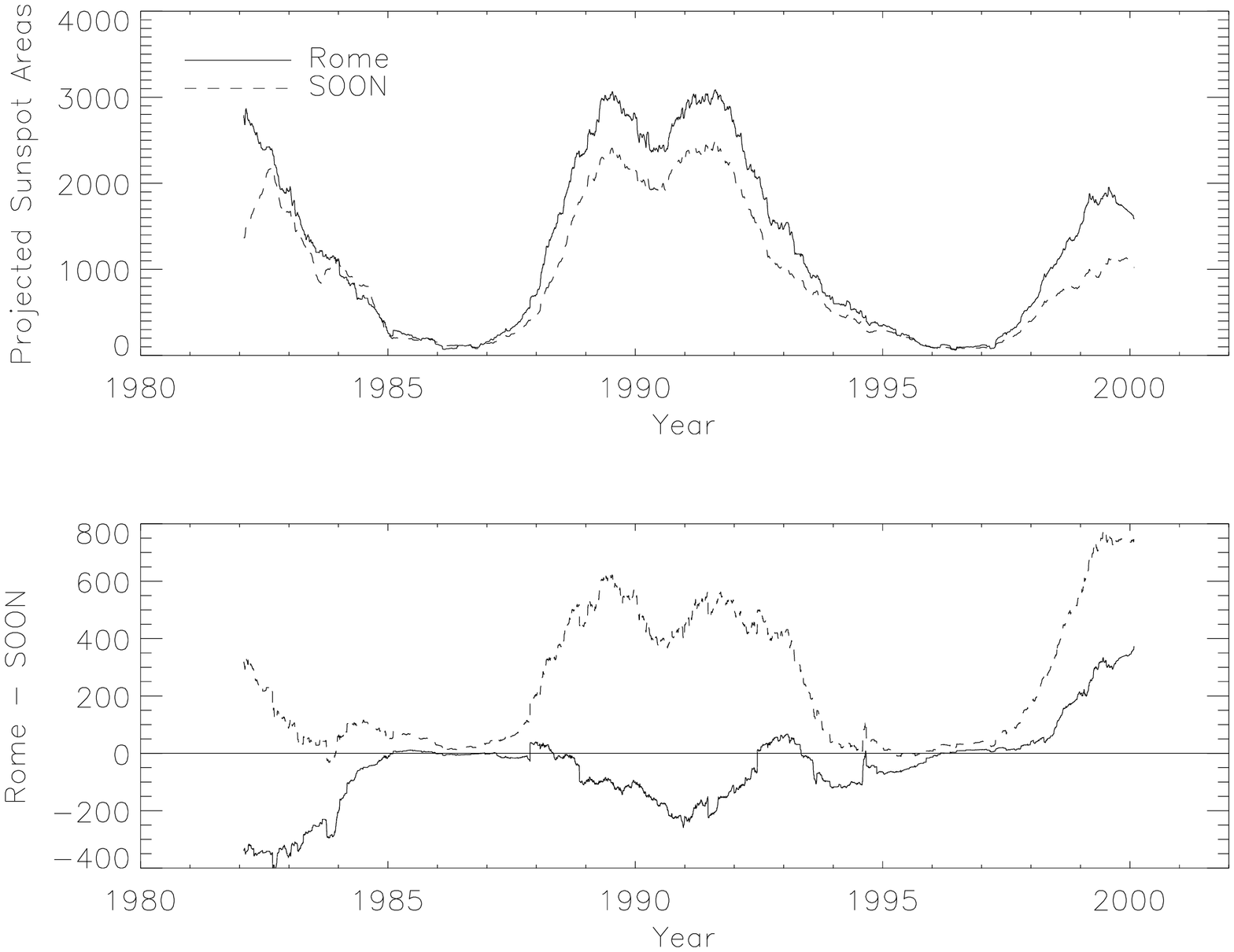}}
\caption{\footnotesize Upper panel: 12-month running means of projected
sunspot areas. Areas are in units of millionths of the solar disc. They
correspond to original values without any calibration. The solid line
represents data from the Rome Observatory, the dashed line represents data from
SOON. Lower panel: The solid line represents the 12-month running
mean of the difference between data from these two observatories after
calibrating the data. The dashed line is the difference between the
original data from both observatories. } \label{fig7}
\end{center}
\end{figure}

In Fig.~\ref{fig7} we show the comparison between data from SOON and Rome, which 
overlap for a long
period of time. A 12-month running mean of the original data vs time (upper 
panel) as well as the
difference between the data from the two observatories, for both original and 
calibrated data (lower
panel), are shown.

In Table~\ref{Table5}, values of the calibration factors for
projected sunspot areas and for areas corrected for foreshortening
obtained using this technique are listed. The corresponding values
for $\mathcal M$ are also tabulated. In all cases, these factors are
lower than the ones found as explained in Section~\ref{analysis} and
Appendix~\ref{offset}. Note, however that if we first form (monthly
or yearly) running means of $A^{bas}$ and $A^{aux}$ before
minimizing $\mathcal M$ we obtain calibration factors much closer to
those listed in Table~\ref{Table4}. This has got to do with the fact
that outliers are given a much smaller weight when forming running
means than if taking the squared difference between daily data.

This technique differs from the one discussed in Section~\ref{analysis} and 
Appendix~\ref{offset} in
that here the same weight is given to maximum and minimum phases of solar cycle. 
It can be seen from
Fig.~\ref{fig7}, that after calibration the difference between both data sets is 
very close to zero
during activity minimum. However, during times of high solar activity this 
calibration technique does
not give as accurate results.

As mentioned before, one of the most important applications of sunspot areas 
data sets is irradiance
reconstruction. So we intend to produce a homogeneus and as complete as possible 
time series of sunspot
areas that can be used in irradiance models to describe adequately the 
variations. Since sunspot
contribution to these variations is most important during times of high 
activity, a method giving larger weight to periods of high activity (large spot areas) should provide a more appropriate calibration factor. For this reason, we use the factors obtained with the method explained in Section~\ref{analysis} and Appendix~\ref{offset} as the default.
Note, however, that in almost all cases factors 
obtained by the two methods
agree within the given uncertainties (even if Eq. \ref{eqb} is applied to daily 
data, without first
forming running means).

%=================================
% TABLES
%=================================

\begin{table*}%[p!]%[t!]
\caption{Data provided by the different observatories} \label{Table1}
\begin{center}
\begin{tabular}{lcccccccc}
\hline
\\
Observatory & Observation Period & Observing technique & Coverage  &
Min. area reported\\
& & &[\%] & [ppm of solar hemisphere]\\
\hline
\\
RGO                &$ 1874 $--$ 1976 $ & photographic plates & 98 & 1\\
Russia             &$ 1968 $--$ 1991 $ & photographic plates & 96 & 1\\
SOON               &$ 1981 $--$ 2008 $ & drawings            & 98 & 10\\
Rome               &$ 1958 $--$ 1999 $ & photographic plates & 50 & 2\\
Catania            &$ 1978 $--$ 1987 $ & drawings            & 81 & 3\\
Yunnan             &$ 1981 $--$ 1992 $ & photographic plates & 81 & 2\\
\hline
\end{tabular}
\end{center}
\end{table*}

\begin{table*}[h!]%[p!]%[h!]
\caption{Calibration factors for the different observatories} \label{Table4}
\begin{center}
\begin{tabular}{lccccccc}
\hline
\\
 Obs.~1 & Obs.~2 & Overlap & Calibration & Calibration & Correlation & 
Correlation \\
        &      & Period     & Factor      & Factor      & Coefficient & 
Coefficient \\
        &      &      & PA          &  CA         &  PA         &  CA\\
\hline
\\

RGO & Russia & $1968-1976$ & $ 1.019 \pm 0.067$ & $ 1.028 \pm
0.083$ & $0.974$ & $0.961$\\
Russia & SOON& $ 1982-1991$ & $ 1.402 \pm 0.131$ & $ 1.448
\pm 0.148$ & $0.953$ & $0.942$\\
RGO & Rome & $1958-1976$ & $1.095 \pm 0.086$ & $1.097 \pm
0.084$ & $0.969$ & $0.963$ \\
Russia & Rome & $1968-1990$ & $1.169 \pm 0.058$ & $1.227 \pm
0.107$ & $0.947$ & $0.907$\\
SOON & Rome & $1982-1999$ & $0.791 \pm 0.105$ &
$0.846 \pm 0.138$ & $0.946$ & $0.902$\\
Russia & Yunnan & $ 1968-1990$ & $ 1.321 \pm 0.215$ & $ 1.365 \pm
0.242$ & $0.959$ & $0.947$ \\
SOON & Yunnan & $1982-1992$& $0.913 \pm 0.113$ & $0.907 \pm
0.131$ & $0.948$ & $0.955$\\
Russia & Catania & $ 1978-1987$ & $ 1.236 \pm 0.052$ & $1.226 \pm
0.059$ & $0.959$ & $0.909$\\
SOON & Catania & $ 1982-1987$ & $0.948 \pm 0.042$ & $0.925
\pm 0.097$ & $0.967$ & $0.949$\\
\hline
RGO & SOON    & via Russia & $1.429 \pm 0.163$ & $1.489 \pm 0.194$ & - & - \\
RGO & Catania & via Russia & $1.240 \pm 0.099$ & $1.234 \pm 0.119$ & - & - \\
RGO & Yunnan  & via Russia & $1.346 \pm 0.237$ & $1.403 \pm 0.273$ & - & - \\
\hline
\end{tabular}
\end{center}
\end{table*}

\begin{table*}[h!]%[p!]%[h!]
\caption{Calibration factors for projected and corrected sunspot areas measured 
by
different observatories obtained by minimizing $\mathcal M$ (see Appendix \ref{otromet})}
\label{Table5}
\begin{center}
\begin{tabular}{lccccc}
\hline
\\
Obs.~1 & Obs.~2 & Calibration & Calibration & $\mathcal M$ & $\mathcal M$ \\
        &         & Factor      & Factor      &  PA &  CA\\
        &         & PA          &  CA         &          &  \\
\hline
\\
RGO    & Russia   & $1.014$  & $ 1.012 $ & $ 68.9$ & $94.6$\\ 
Russia & SOON     & $1.352$  & $ 1.399 $ & $136.1$ & $242.7$\\ 
RGO    & Rome     & $1.083$  & $ 1.073 $ & $ 61.3$ & $70.3$\\
Russia & Rome     & $1.121$  & $ 1.106 $ & $220.5$ & $185.9$\\ 
SOON   & Rome     & $0.764$  & $ 0.733 $ & $215.8$ & $157.6$ \\ 
Russia & Yunnan   & $1.279$  & $ 1.326 $ & $111.5$ & $130.6$\\
SOON   & Yunnan   & $0.895$  & $ 0.882 $ & $ 75.6$ & $87.9$\\ 
Russia & Catania  & $1.205$  & $ 1.197 $ & $202.1$ & $164.5$\\ 
SOON   & Catania  & $0.924$  & $ 0.853 $ & $197.5$ & $244.5$\\ 
\hline
\end{tabular}
\end{center}
\end{table*}

\acknowledgements{This work was supported by the \textit{Deutsche
Forschungsgemeinschaft, DFG} project number SO 711/1-1. We thank M. Lockwood
for his encouragement and critical discussions and P. Foukal for providing
the plage and enhanced network areas data set as well for the information and
techniques used by him.}

\bibliographystyle{agu}
% \bibliography{biblio}

\end{article}
\end{document}